\documentclass[twoside,final]{pro12}

\usepackage[ansinew]{inputenc}
\usepackage{amsmath}
\usepackage{graphicx}
\usepackage{array}
\usepackage{SIunits}
\usepackage{hyperref}
\usepackage{lscape}
\usepackage{multirow}
\usepackage{float}
\usepackage{natbib}
\usepackage{hyperref}
\defcitealias{Turner2021}{T21}
\usepackage{ragged2e}
\usepackage{threeparttable}
\usepackage{xcolor}
\usepackage{subcaption}
\usepackage{calc}
\usepackage{lineno}
\usepackage{enumerate}

\newenvironment{correction}{}

\head{Turner et al.}{Follow-up NenuFAR observations of the $\tau$~Boo}

\begin{document}

\title{Follow-up radio observations of the $\tau$~Bo\"{o}tis exoplanetary system: Preliminary results from NenuFAR}	

\author{J.D. Turner\adress{\textsl Department of Astronomy and Carl Sagan Institute, Cornell University, Ithaca, NY, USA} $^{,}$\adress{\textsl NHFP Sagan Fellow}
$^{*}$, 
P. Zarka\adress{\textsl Observatoire Radioastronomique de Nan\c{c}ay (ORN), Observatoire de Paris, PSL Research University, CNRS, Univ. Orl\'{e}ans, OSUC, 18330 Nan\c{c}ay, France} $^{,}$\adress{\textsl LESIA, Observatoire de Paris, CNRS, PSL, Meudon, France}
, 
J-M. Grie{\ss}meier$^{3,}$\adress{Laboratoire de Physique et Chimie de l'Environnement et de l'Espace (LPC2E) Universit\'{e} d'Orl\'{e}ans/CNRS, Orl\'{e}ans, France}
, \\ 
E. Mauduit$^4$, 
L. Lamy$^{3,4,}$\adress{Aix Marseille Universit\'e, CNRS, CNES, LAM, Marseille, France}
, 
J.N. Girard$^4$,\\
T. Kimura\adress{Department of Physics, Tokyo University of Science, Tokyo, Japan}
,
 B. Cecconi$^4$ 
, 
L.V.E. Koopmans\adress{Kapteyn Astronomical Institute, University of Groningen, Groningen, The Netherlands}
\\
\footnotesize $^*$Corresponding author: \href{mailto:jaketurner@cornell.edu}{jaketurner@cornell.edu} \normalsize
}

\maketitle

\begin{abstract}
Studying the magnetic fields of exoplanets will provide valuable information about their interior structures, atmospheric properties (escape and dynamics), and potential habitability. One of the most promising methods to detect exoplanetary magnetic fields is to study their auroral radio emission. However, there are no confirmed detections of an exoplanet in the radio despite decades of searching. Recently, \citet{Turner2021} reported a tentative detection of circularly polarized bursty emission from the $\tau$~Bo\"{o}tis ($\tau$~Boo) exoplanetary system using LOFAR low-frequency beamformed observations. The likely source of this emission was presumed to be from the $\tau$ Bootis planetary system and a possible explanation is radio emission from the exoplanet $\tau$~Boo~b, produced via the cyclotron maser mechanism. Assuming the emission is from the planet, \citet{Turner2021} found that the derived planetary magnetic field is compatible with theoretical predictions. The need to confirm this tentative detection is critical as a conclusive detection would have broad implications for exoplanetary science. In this study, we performed a follow-up campaign on the $\tau$~Boo system using the newly commissioned NenuFAR low-frequency telescope in 2020. We do not detect any \correction{bursty} emission in the NenuFAR observations. There are many different degenerate explanations for our non-detection. For example, the original bursty signal may have been caused by an unknown instrumental systematic. Alternatively, the planetary emission from $\tau$~Boo~b is variable. As planetary radio emission is triggered by the interaction of the planetary magnetosphere with the magnetized stellar wind, the expected intensity of the planetary radio emission varies greatly with stellar rotation and along the stellar magnetic cycle. More observations are needed to fully \correction{understand} the mystery of the possible variability of the $\tau$~Boo~b radio emission. 
\end{abstract}

\section{Introduction}

The search for auroral radio emissions from exoplanets has been ongoing for many decades. In analogy with the magnetized Solar System planets and moons, these radio emissions are expected to be produced via the \correction{cyclotron maser instability} (CMI) mechanism (\citealt{Wu1979,Zarka1998,Treumann2006}) and be highly circularly polarized, beamed, and time-variable (e.g., \citealt{Zarka1998,Zarka2004}). CMI emission is produced at the local electron cyclotron frequency ($\omega_{c}$) in the source region and the maximum gyrofrequency $\nu_{g}$ is determined by the maximum magnetic field $B_{p}$ near the planetary surface, as $\nu_{g} [MHz] =  2.8 \times B_{p} [G]$, \correction{where the frequency is measured in MHz and the magnetic field is measured in Gauss.} Therefore, radio observations can be used to probe \correction{the magnetic fields of exoplanets} (e.g. \citealt{Farrell1999,Zarka2001,Zarka2007}). 
 
Low-frequency radio observations are indeed among the most promising methods to detect their magnetic fields (\citealt{G2015}) as many of the other techniques are prone to false-positives (e.g. \citealt{G2015,Alexander2016, Turner2016a,Route2019,Strugarek2022}). Measuring the magnetic field of an exoplanet will give valuable constraints on its interior structure, atmospheric escape, and star-planet interactions (\citealt{Zarka2015SKA,Lazio2016,Griessmeier2018haex,Zarka2018haex,Lazio2019}). Also, atmospheric dynamics may be altered by the presence of a planetary magnetic field (e.g., \citealt{Perna2010a,Rauscher2013, Hindle2021}) and Ohmic dissipation may contribute to the anomalously large radii of hot Jupiters (e.g. \citealt{Perna2010a,Knierim2022}). Finally, a magnetic field might be one of the many properties needed on Earth-like exoplanets to sustain their habitability (e.g., \citealt{Gr2005,Lammer2009,Meadows2018haex,McIntyre2019}). 

 Much work has been done to search for exoplanet radio emission for many decades and a thorough overview of the theory and observations can be found in the reviews by \citet{Zarka2015SKA}, \citet{G2015}, and \citet{Griessmeier17PREVIII}. Following several seminal papers (\citealt{Zarka1997pre4,Farrell1999,Zarka2001,Zarka2007}), 
a large body of theoretical work has been published (e.g., \citealt{Lazio2004,Griessmeier05AA,Gr2007,Hess2011,Nichols2012,Weber2017pre8,Lynch2018,Ashtari2022,Griessmeier23PRE9}). In parallel to these theoretical studies, a number of ground-based observations have been conducted to find \correction{exoplanet} radio emissions and most of them resulted in unambiguous non-detections (e.g. \citealt{Yantis1977,Winglee1986,Zarka1997pre4,Bastian2000,Hallinan2013,Turner2017pre8,Lynch2018,Cendes2022,Narang2023}). There are many degenerate reasons for the non-detections that are discussed in \citet{Griessmeier17PREVIII} and references \correction{therein}. Currently, there are a few tentative detections (\citealt{Lecavelier2013,Sirothia2014,Vasylieva2015,Bastian2018}) but none of these have been confirmed by follow-up observations. By contrast, progress has been made in detecting free-floating planets near the brown dwarf boundary (\citealt{Kao2016,Kao2018}) \correction{and stellar emission, which could potentially originate from star-planet interactions (e.g., \citealt{Vedantham2020,Callingham2021,Perez2021,Pineda2023,Trigilio2023}; \citealt[and references within]{BlancoPozo2023}).} 

Recently, \citet[hereafter T21]{Turner2021} tentatively detected circularly polarized bursty and slow emission from the $\tau$~Bo\"{o}tis ($\tau$~Boo) exoplanetary system using LOFAR beamformed observations. The \correction{slow and bursty} emissions were only seen around a phase (relative to the periastron) of 0.65 and 0.8, respectively (see Figure \ref{fig:orbits}). The slow emission was detected between 21--30 MHz, while the bursty emission was detected between 15--21 MHz. 
For the slow emission the signal could neither be confirmed with certainty nor could it be fully refuted; however, for 
the bursty emission \citetalias{Turner2021} found no potential cause of a false positive. \citetalias{Turner2021} determined that the signal originated from the $\tau$~Boo system and that CMI radio emission from the exoplanet $\tau$~Boo~b is the most likely explanation. Assuming their radio signals originated from the planet, \citetalias{Turner2021} derived constraints on the magnetic field of $\tau$~Boo b that are consistent with theoretical predictions (\citealt{Gr2007,Griessmeier17PREVIII}). More recent calculations with a new implementation of such predictions also find compatible values for the planetary magnetic moment and flux density (\citealt{Ashtari2022,Mauduit23PRE9}) \correction{and the phases that have been observed to have emission (\citealt{Ashtari2022}).}
\correction{On the other hand, the derived constraints on the magnetic field of $\tau$~Boo b by \citetalias{Turner2021} are in conflict with the field strength derived by \citet{Cauley2019} via optical star-planet interaction (SPI) observations. However, several new studies have shown that the original SPI signatures are not statistically robust and need to be re-evaluated(\citealt{Route2019,Strugarek2022}). Therefore, more work is needed before we can compare the SPI and radio observations.} 
More follow-up observations were highly advocated by \citetalias{Turner2021} to confirm their tentative detections and to search for periodicity in the signal. 

Motivated by the tentative detection of radio emission from $\tau$~Boo, we performed a large follow-up campaign at low-frequencies. This campaign was coordinated between four different radio telescopes including LOFAR and NenuFAR. In this paper, we only present the NenuFAR data of $\tau$~Boo. The LOFAR data will be presented elsewhere.

\section{Observations}
 
All observations in this paper were taken with NenuFAR (New Extension in Nan\c{c}ay Upgrading LOFAR) (\citealt{Zarka2020}), a large phased array located at the Nan\c{c}ay Radio observatory.
The data streams from all mini-arrays (each composed of 19 individual antennas) were coherently summed together during the observations. All beamformed observations are obtained with the UnDySPuTeD receiver (described in \citealt{Bondonneau2021}), which provides the full Stokes parameters. As part of the Early Science Phase (2019--2022), NenuFAR observed a large sample of exoplanets and radio-active stars through a dedicated Key Project. The $\tau$~Boo observations presented in the following were taken as part of this Key Project. 

\begin{figure}[!htb]
\centering
\includegraphics[width=0.8\textwidth]{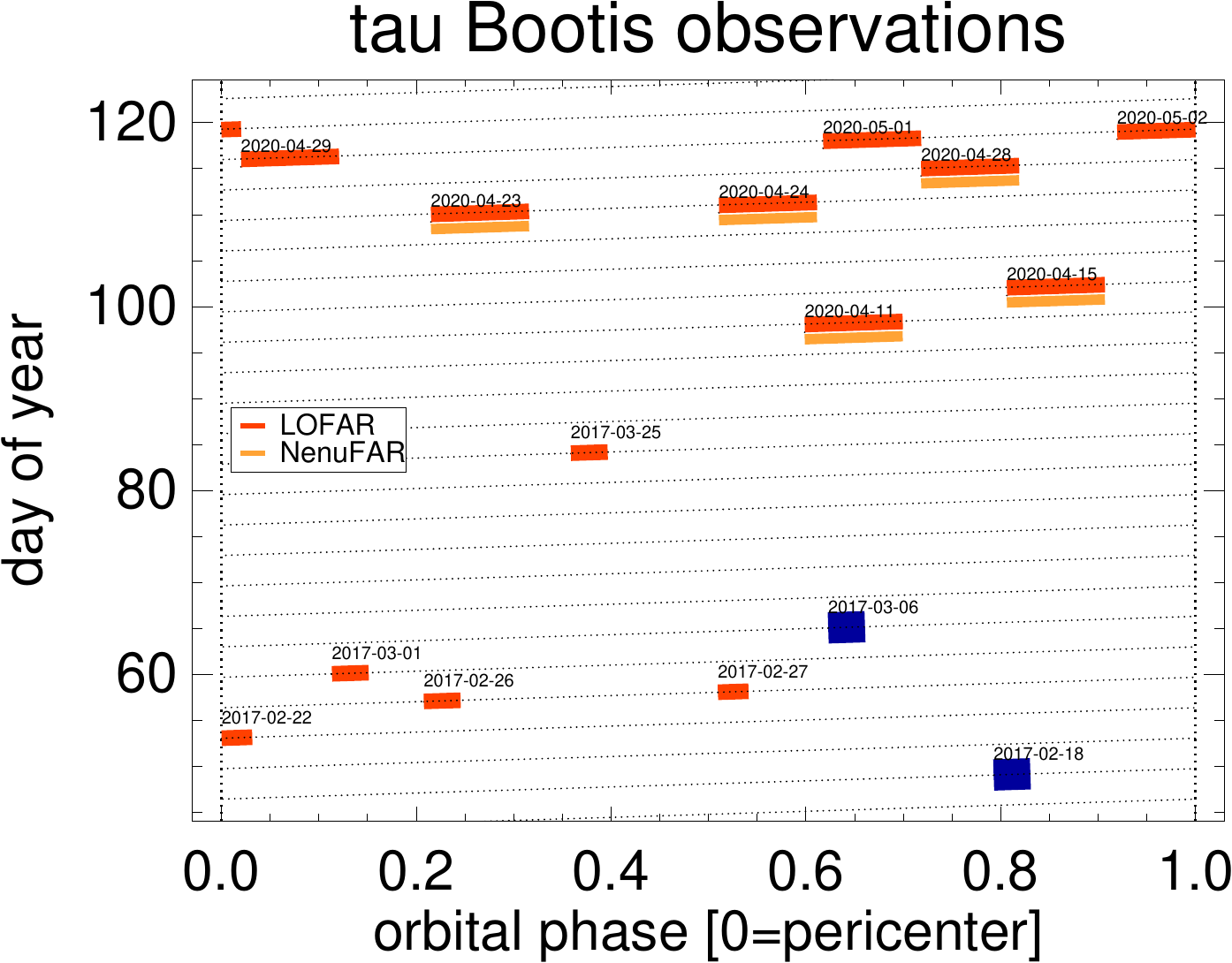}
\caption{Orbital phase coverage for the observations of $\tau$~Boo~b. The orbital phase was calculated using the ephemeris from \citet{Wang2011} and is relative to periastron of $\tau$~Boo~b. The LOFAR and NenuFAR observations are labeled in red and orange, respectively.  The original LOFAR observations were taken between February 18, 2017 and March 25, 2017. The $\tau$~Boo LOFAR observations with the tentative detections are displayed as large dark-blue rectangles. The tentative detection of burst emission was seen on February 18, 2017 and the slow emission was seen on March 6, 2017. An arbitrary number of days was placed between the 2020 and 2017 observations for clarity.}
\label{fig:orbits}
\end{figure}

We observed $\tau$~Boo for 64 hours with NenuFAR from 14-52 MHz in beamformed mode with full Stokes in April and May 2020. In this study, we will only concentrate on the \correction{Stokes V (circularly polarized flux)} data since auroral radio emissions are expected to be circularly polarized (\citealt{Zarka1998,Zarka2004}), \citet{Turner2019} showed that \correction{Stokes V} data is an order of magnitude more sensitive than \correction{Stokes I (total intensity)} data for the detection of attenuated Jovian radio bursts, 
and the \citetalias{Turner2021} tentative detection was in \correction{Stokes V}. The observations consist of an on-target beam (``ON-beam'') and three beams pointing to a nearby location in the sky (``OFF-beam 1'', ``OFF-beam 2'', and ``OFF-beam 3''). As in \citet{Turner2017pre8,Turner2019,Turner2021}, we use the OFF-beams to characterize any terrestrial ionospheric fluctuations, RFI (radio frequency interference), and instrumental systematics in the data. In the final analysis, the OFF-beams are used to remove all these signals from the ON-beam. Therefore, any remaining signal in the ON-beam should be astrophysical in nature. 
The setup of the observations can be found in Table \ref{tb:setup} \correction{and the positions of the ON- and OFF-beams can be found in Table \ref{tb:beam}. The location of the OFF-beams were chosen such that they are located at an angular distance of 2.8$\pm$0.4$^\circ$ (the second null location of NenuFAR's PSF; \citealt{Zarka2020}) from $\tau$~Boo and that they are devoid of point radio sources from the LOFAR Multifrequency Snapshot Sky Survey (MSSS; \citealt{Heald2015}) with a flux $\ge$ 300 mJy.} Our observations used the 53 mini-arrays available during the Early Science Phase. We performed the observations at night to avoid strong contamination by RFI. The exact dates and times of each observation can be found in Table \ref{tb:obs} \correction{and the orbital phases\footnote{\correction{With the error bars of \citet{Wang2011}, the orbital phase shifts by $\sim$10$\%$ over a period of 100 years.}} covered by the observations can be found in Figure \ref{fig:orbits}.} Before each observation, 
we observed the pulsar B0809+74 for 17 mins using the exact same setup. We use the pulsar observations \correction{to ensure the reliability of the data reduction pipeline (e.g. adequate RFI masking), compare the ON and OFF beams, and to search for systematics effects in the data.} In \citetalias{Turner2021}, similar B0809+74 observations were used to understand and rule out certain systematics as the cause of the detected signals (see their Appendix G). Three of the observations (April 29, May 1, and May 2) experienced technical difficulties resulting in only one OFF beam being recorded to disk. Due to this problem, these observations were not used in our analysis. Therefore, the useful set of observations covered 50$\%$ of $\tau$~Boo's orbit. This is twice the coverage of the \citetalias{Turner2021} LOFAR observations. Most importantly, the NenuFAR observations cover the same orbital phases as the tentative detections in \citetalias{Turner2021} (Figure \ref{fig:orbits}). 

\begin{table}[!htb]
\centering
\caption{Setup of NenuFAR observations}
\begin{tabular}{ccc}
\hline 
\hline
Parameter & Value  & Units \\ 
\hline
Beams                           & ON, 3 OFF     \\
 Stokes parameters              & IQUV          \\
 Number of mini-arrays          & 53              \\
Antennas per mini-array         & 19\\ 
Total number of antennas        & 1007 \\
 Minimum Frequency              & 14.8          & MHz\\
 Maximum Frequency              & 52.1          & MHz\\
 Frequency Resolution           & 3.05          & kHz      \\
 Time Resolution                & 10          & msec     \\
 Subbands                       & 192               \\
 Channels per Subband           & 64          &   \\
 Beam diameter$^{a}$                  & 2.39         & $^{\circ}$\\ 
\hline
\end{tabular}
\begin{tablenotes}
\centering
\item $^{a}$Calculated at 15 MHz (\citealt{Zarka2020})
\end{tablenotes}
\label{tb:setup}
\end{table}

\begin{table}[!htb]
\centering
\caption{\correction{Beam coordinates used for the observations} } 
\begin{tabular}{cccc}
\hline 
\hline
Beam   & RA (J2000)  & DEC (J2000)  & Distance \\
       &  (h:m:s)    & ($^\circ$:':") & ($\degree$) \\
\hline 
ON      & 13:47:17      & 17:27:22      & --- \\
OFF 1   & 13:49:04	    &20:20:37	   &2.9    \\
OFF 2   & 13:38:50	    &15:23:37	   &2.9 \\
OFF 3   & 13:55:48	    &15:21:22	   & 2.9 \\
\hline
\end{tabular}
\begin{tablenotes}
\centering
\item Note: Column 1: beam name. Column 2: right ascension (RA) of the beam. Column 3: declination (DEC) of the beam. Column 4: distance of the beam from the ON-beam.
\end{tablenotes}
\label{tb:beam}
\end{table}

\begin{table}[!htb]
\centering
\caption{Summary of NenuFAR observations}
\begin{tabular}{ccccc}
\hline 
\hline
Obs $\#$ &Start Date      & Time          & Duration    &  Phase  \\ 
  & (UT)            & (UT)          & (hrs) \\
\hline
\multicolumn{5}{c}{$\tau$~Bo\"{o}tis [64 hrs]}\\
\hline
1&April 11, 2020  & 20:00-04:00    & 8            & 0.60--0.70\\
2&April 15, 2020  &20:00--04:00     & 8            & 0.81--0.91\\
3&April 23, 2020  &19:30--03:30    & 8            & 0.22--0.32 \\
4&April 24, 2020  &19:00--03:00     & 8            & 0.51--0.61 \\ 
5&April 28, 2020  &19:00--03:00     & 8            & 0.72--0.82\\
6$^{a}$&April 29, 2020  &19:00--03:00     & 8            & 0.02--0.12     \\
7$^{a}$&May 1, 2020     &18:30--02:30      &8           & 0.62--0.72      \\
8$^{a}$&May 2, 2020     &18:30--02:30      & 8          & 0.92--0.02     \\
\hline
\multicolumn{5}{c}{B0809+74 [2.27 hrs]}\\
\hline
1&April 11, 2020  & 19:41--19:58 &0.28 & --  \\
2&April 15, 2020  &19:42--19:59 & 0.28 & --\\
3&April 23, 2020  &19:11--19:28 & 0.28 & --\\
4&April 24, 2020  &18:41--18:58 &0.28& --\\
5&April 28, 2020  &18:41--18:58 &0.28& -- \\
6$^{a}$&April 29, 2020  &18:41--18:58 &0.28& -- \\
7$^{a}$&May 1, 2020     &18:11--18:28 &0.28& -- \\
8$^{a}$&May 2, 2020     &18:11--18:28 &0.28& -- \\
\hline
\end{tabular}
\begin{tablenotes}
\centering
\item $^{a}$Only 1 OFF beam was recorded
\end{tablenotes}
\label{tb:obs}
\end{table}

\section{Data Reduction}

The raw beamformed exoplanet observations from NenuFAR are produced by coherently summing all the signals from each dipole antenna. They represent 125 GB / hour stored at the Nan\c{c}ay data center. The reduction pipeline applied to our data consisted in computing the 4 Stokes parameters (I, Q, U, and V), and applying RFI mitigation and correction of the time variations of the instrumental gain, before integrating the measurements on 250 ms $\times$ 48 kHz bins (i.e. 24 $\times$ 16 raw bins). The purpose of the RFI mitigation is to prevent polluted 3.05 $\times$ 10 ms bins to be included in the integrated 250 ms $\times$ 48 kHz bins, thus it is restricted to scales smaller than 250 ms in time and 48 kHz in frequency. RFI at larger scales are not eliminated at the reduction stage, and must be flagged in the post-processing stage. For the 2020 data, only the intensity (I), circular polarization (V) and linear polarization ($L=(Q^2+U^2)^{1/2}$ were stored. The reduced data, flag masks, time and frequency ranges, and housekeeping data are stored as standard fits files, of volume $\sim$400 times smaller than the raw data, thus easy to post-process. This data is labeled as level 2 data (L2) whereas the level 1 data (L1) is the raw data. The raw data are then erased. 

\begin{figure}[!thb]
\centering
\begin{flushleft} \textbf{(A)} \end{flushleft}    
\includegraphics[page=3,width=\textwidth]{TauBoo_Dynspec_L2-crop.pdf}
 \begin{flushleft} \textbf{(B)} \end{flushleft}  
\includegraphics[page=1,width=\textwidth]{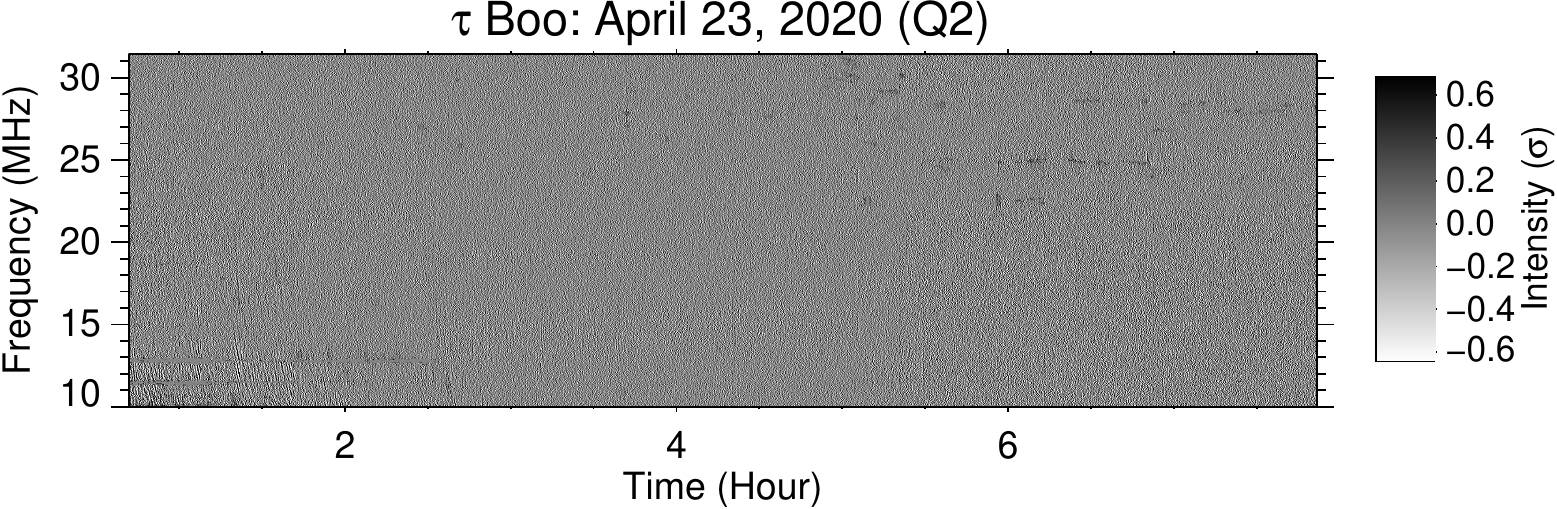}  \\ 
  \caption{The dynamic spectra in \correction{Stokes V} for the April 23, 2020 $\tau$~Boo observation with NenuFAR. \textbf{(A.)} Example of a pre-processed (L2) dynamic spectra. The intensity in the dynamic spectrum is in arbitrary units. \correction{\textbf{(B.)} Example of a high-pass filtered dynamic spectra.} The dynamic spectrum is normalized by the standard deviation of the observations.} 
  \label{fig:Dynspec_all}
\end{figure}

An example dynamic spectra of the L2 data for $\tau$~Boo can be found in Figure \ref{fig:Dynspec_all}A. In this figure, lots of systematics are still clearly visible in the data. The source of many of the features is not yet well understood but several have a known origin. For example, the analog beams of NenuFAR are re-pointed every 6 minutes to ensure accurate source tracking. These re-pointings are clearly visible in the data as the baseline changes after every new pointing. The blobs above 30 MHz are caused by bright-sources in the grating lobes of NenuFAR. 

Next, the L2 data were processed using the \texttt{BOREALIS} (BeamfOrmed Radio Emission AnaLysIS) pipeline (\citealt{Turner2017pre8,Turner2019,Turner2021}). We performed RFI mitigation using a similar setup as in \citetalias{Turner2021}. Due to the complexity of the NenuFAR data (e.g. six-minute pointing jumps), we did not correct for the time-frequency background of the observations. This will be done in future work. This limitation is the reason why we only search for bursty emission in this study. After RFI mitigation we re-binned the data to a time and frequency resolution of 1 second and 45 kHz, respectively. This rebinned data was then fed into the post-processing part of \texttt{BOREALIS}. 

\begin{figure}[!thb]
\centering
\includegraphics[page=1,width=0.6\textwidth]{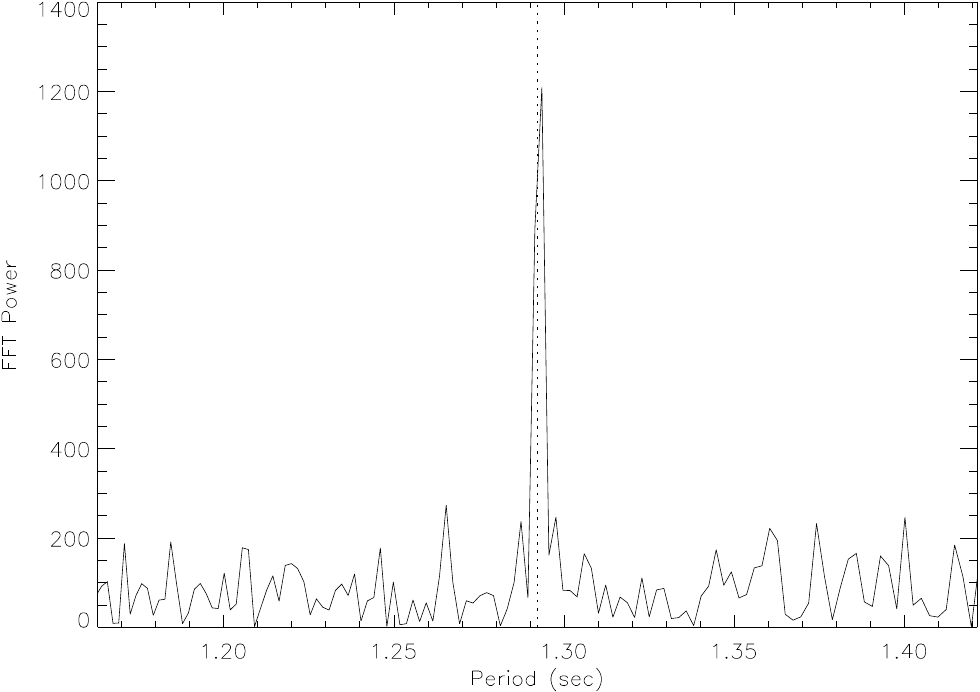}  \\ 
  \caption{FFT of the pulsar B0809+74 for the NenuFAR data taken on taken on April 23, 2020. The known period is denoted as the dashed line and the pulsar is detected with a signal-to-noise of $\sim$6 around this period. } 
  \label{fig:FFT}
\end{figure}

In order to determine the reliability of the NenuFAR data, we processed the pulsar (B0809+74) through the \texttt{BOREALIS} pipeline. In order to detect B0809+74 we perform an FFT on the RFI masked, de-dispersed (the data is de-disperse using the pulsar's known dispersion measure), and rebinned data (we only rebinned the data in frequency). The FFT was computed using data from 40-50 MHz. Figure \ref{fig:FFT} shows an FFT of B0809+74 using NenuFAR data taken on April 23, 2020. The pulsar is detected with a signal-to-noise of $\sim$6 at its known period. 
\correction{Based on this result, we are confident that the pipeline masks the RFI adequately and will allow for the search for bursty signals from the exoplanet.}

In the post-processing, we only searched for burst emission. A very detailed description of the burst emission observables can be found in \citet{Turner2019} and \citetalias{Turner2021} and we describe them briefly below. The post-processing is performed on the absolute value of the corrected \correction{Stokes V} data as defined in \citet{Turner2019}. For burst emission, we only use the Q2 and Q4a-f observable quantities. The Q2 observable consists of one time series per beam (ON, OFF1, OFF2, and OFF3). It is obtained by high-pass filtering the dynamic spectrum. An example of a high-pass filtered dynamic spectrum for $\tau$~Boo can be found in Figure \ref{fig:Dynspec_all}B. To search for faint emission, the high-pass filtered dynamic spectrum is integrated over a large frequency range (e.g., bin sizes of 10 MHz) to produce a time series. Q2 can be represented by a ``scatter plot'' comparing a pair of beams (e.g., the ON and one of the OFF beams) and is designed to find bursty emission (with time scales $<\sim$ 1 minute). The quantities Q4a to Q4f provide statistical measures of the bursts identified by the Q2 quantity. When examining Q4a-f, the ON and OFF time series are compared to each other; for this, we introduce the difference curve Q4f$_{\text{Diff}}$=Q4f(ON)-Q4f(OFF). We then plot this curve against a reference curve computed from 10000 draws of purely Gaussian noise. 

The post-processing was performed separately over 4 different frequency ranges (15 - 21 MHz, 15-52 MHz, 15-33 MHz, and 33-52 MHz). \correction{All other post-processing parameters are similar to those of \citetalias{Turner2021}.} In this paper, we search for burst emission in the data that has binned to 1 second and highpass filtered with a smoothing timescale of 10 seconds. Therefore, only emission less than 10 seconds can be searched for in the data. \citetalias{Turner2021} only observed burst emission at this same timescale. 

\section{Results}
\begin{figure*}[!htb]
\centering
  \begin{tabular}{cc}
 \begin{subfigure}[t]{0.48\linewidth}
      \centering
      \captionsetup{justification=raggedright, singlelinecheck=false}
     \caption*{\textbf{(a)}}
      \includegraphics[width=0.90\linewidth,page=1]{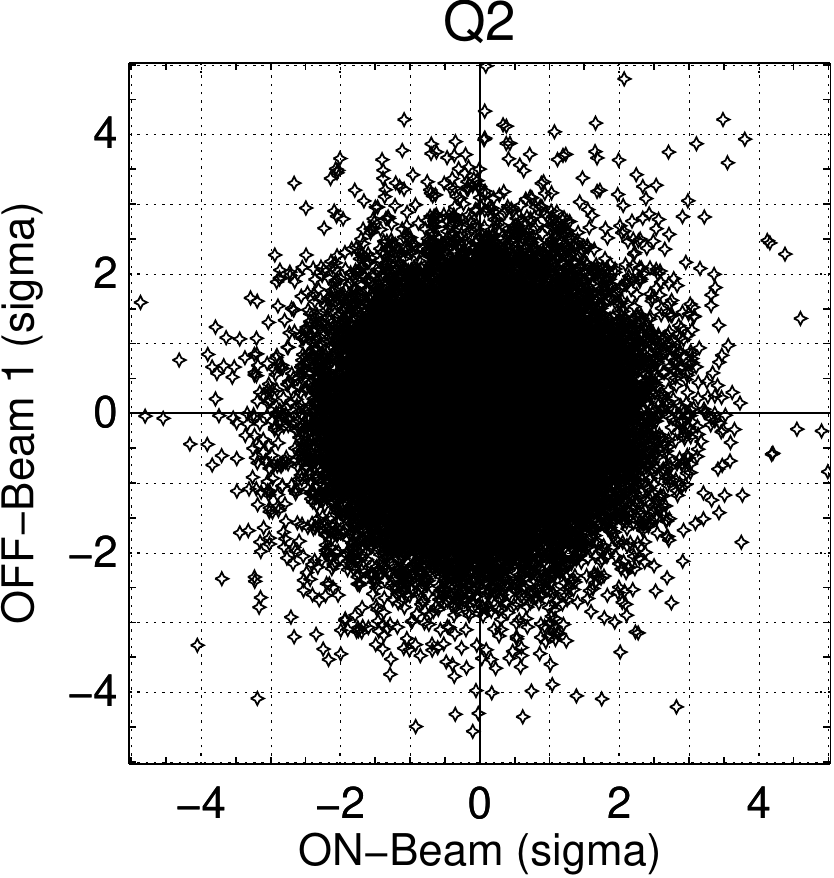}
    \end{subfigure} & 
    \begin{subfigure}[t]{0.48\linewidth}
      \centering
            \captionsetup{justification=raggedright, singlelinecheck=false}
           \caption*{\textbf{(b)}}
      \includegraphics[width=0.90\linewidth,page=2]{20200415_14.6to33.3MHz_Nondetection-crop.pdf}
    \end{subfigure} \\
    \begin{subfigure}[t]{0.48\linewidth}
      \centering
                  \captionsetup{justification=raggedright, singlelinecheck=false}

        \caption*{\textbf{(c)}}
      \includegraphics[width=0.90\linewidth,page=4]{20200415_14.6to33.3MHz_Nondetection-crop.pdf}
     
    \end{subfigure} & 
     \begin{subfigure}[t]{0.48\linewidth}
      \centering
        \captionsetup{justification=raggedright, singlelinecheck=false}
 \caption*{\textbf{(d)}}
      \includegraphics[width=0.90\linewidth,page=6]{20200415_14.6to33.3MHz_Nondetection-crop.pdf}
     
    \end{subfigure} \\
    \begin{subfigure}[t]{0.48\linewidth}
      \centering
                  \captionsetup{justification=raggedright, singlelinecheck=false}
 \caption*{\textbf{(e)}}
      \includegraphics[width=0.90\linewidth,page=8]{20200415_14.6to33.3MHz_Nondetection-crop.pdf}
     
    \end{subfigure} & 
    \begin{subfigure}[t]{0.48\linewidth}
      \centering
                  \captionsetup{justification=raggedright, singlelinecheck=false}
     \caption*{\textbf{(f)}}
      \includegraphics[width=0.90\linewidth,page=10]{20200415_14.6to33.3MHz_Nondetection-crop.pdf}
 
    \end{subfigure}
    \end{tabular}
  \caption{Non-detection of burst emission for $\tau$~Boo in the April 15, 2020 NenuFAR observation between 15-32 MHz in \correction{Stokes V}. 
  \textit{Panel a:} Q2 for the ON-beam vs the OFF-beam 2. 
  \textit{Panel b:} Q2 for the OFF-beam 1 vs the OFF-beam 2.
  \textit{Panel c:} Q4a (number of peaks). \textit{Panel d:} Q4b (power of peaks). \textit{Panel e:} Q4e (peak offset). \textit{Panel f:} Q4f (peak offset). For \textit{panels c} to \textit{f} the black lines are the ON-beam difference with the OFF beam 2 and the red lines are the OFF-beam difference. The dashed lines are statistical limits (1, 2, 3$\sigma$) of the difference between all the Q4 values derived using two different Gaussian distributions (each performed 10000 times). We do not see any excess signal in the ON-beam compared to the OFF-beams. 
}
  \label{fig:Nondetection}
\end{figure*}

We searched for excess signal in the ON-beam both by eye and using the automated search procedure outlined in \citetalias{Turner2021}. We applied the same criteria as in \citealt{Turner2019,Turner2021} for a possible detection. We do not find any burst emission in the NenuFAR observations for any of the probed wavelength ranges. The burst statistics were similar between all ON and OFF beams. An example of the diagnostics for a non-detection can be found in Figure \ref{fig:Nondetection}, based on the April 23, 2020 NenuFAR observation, analyzed between 15-32 MHz. The ON-OFF and OFF-OFF difference curves are very similar to each other and all curves are systematically non-Gaussian. We performed a red-noise analysis on the Q2 time series for the ON and OFF beams using the wavelet technique of \citet{Carter2009} as implemented in the \texttt{EXOMOP} code by \citet{Turner2016b}. The wavelet technique assumes that the time series is an additive combination of noise with Gaussian white noise and red noise (characterized as a power spectral density proportional to $1/f^{\alpha}$). We find that the red and white noise components are nearly equal in magnitude. We will investigate this systematic noise in the data in future work.

\section{Discussion}
We do not detect any burst emission in our NenuFAR observations. This is consistent with the simultaneous LOFAR observations that also did not detect any bursty emission that will be presented elsewhere. The original LOFAR observations \citepalias{Turner2021} only detected burst emission flux during one four-hour observation centered around an orbital phase of 0.8 (Figure \ref{fig:orbits}). All the NenuFAR observations besides April 15, 2020 are consistent with this finding. Based on \citetalias{Turner2021}, one might have expected to detect bursty emission at orbital phase 0.8, i.e. for the NenuFAR observation of April 15, 2020. 

\correction{There are many different degenerate reasons for our non-detection of burst emission from $\tau~$Boo:} 
\begin{enumerate}[(a)]
    \item It is possible that the original 3.2$\sigma$ bursty signal detected by \citetalias{Turner2021} was caused by an unknown instrumental systematic or statistical anomaly. \correction{An extensive systematic study was undertaken by \citetalias{Turner2021} and no plausible cause for a false positive signal could be identified for the bursty emission signature. In the future, a statistical comparison between many LOFAR and NenuFAR datasets may shed light on this issue.} 
    \item An extended, evaporating atmosphere can lead to conditions in the 
planetary magnetosphere 
that doesn't allow the CMI mechanism to operate (the local electron plasma frequency is greater than the local electron cyclotron frequency, $\omega_{p}$ $>$ $\omega_{c}$; \citealt[e.g.][]{Griessmeier23PRE9}). 
However, for a planet as massive as $\tau~$Boo~b, this effect is not expected \citep{Weber2018}.
On the other hand, if the planetary magnetic field is as low as estimated by \citet{Reiners2010}, the condition for the CMI to operate is only marginally fulfilled \citep[][Figure 5]{Weber2018}, and temporal variations could, in principle, lead to a situation where the emission does not operate continuously. \correction{It is worth noting that \citet{Kao2016,Kao2018} showed that the \citet{Reiners2010} magnetic field predictions may not be applicable for all brown dwarfs. However, it is currently unclear if the contradiction found by \citet{Kao2016,Kao2018} extends to hot Jupiters such as $\tau$~Boo~b. More work is needed to understand this limitation.} 
\item The observations could have been taken at an unfavorable rotational phase of the star. Planetary radio emission is triggered by the interaction of the planetary magnetosphere with the magnetized stellar wind (\citealt{Zarka2001,Zarka2018haex}). For this reason, the expected intensity of the planetary radio emission differs greatly with stellar rotation (e.g. \citealt{Griessmeier05AA,Griessmeier07PSS,Fares2010,Vidotto2012,See2015,Strugarek2022}). \citet{See2015} showed that the expected radio flux of $\tau~$Boo b can disappear entirely for certain stellar rotation phases. \correction{We cannot rule out variability in the radio signal due to observing at different stellar rotation phases (no stellar differential rotation constraints exist during our 2020 observing campaign).}

\item The observations could have been taken at an unfavorable time in the stellar magnetic cycle. The host star $\tau~$Boo is known to undergo a rapid magnetic cycle of 120 days (\citealt{Jeffers2018}). Similarly to the stellar rotation, this is expected to modulate the planetary radio emission (e.g. \citealt{Fares2009,Vidotto2012,See2015,Elekes2023}). \citet{See2015} also found that the expected radio flux of $\tau~$Boo b (at optimal stellar longitude) can vary by a factor of $\sim$5 along the stellar magnetic cycle. Most recently, \citet{Elekes2023} showed that a magnetic polarity reversal of the host star towards an anti-aligned field configuration would result in the planetary radio emission being below the detection threshold of NenuFAR. However, no stellar magnetic field data of $\tau$~Boo exists during our 2020 observing campaign. \correction{Therefore, we cannot rule out variability in the radio signal due to observing at different moments of the stellar magnetic cycle.}  
\item \correction{Unlike Jupiter's decametric emission (\citealt{Zarka1998}), the planetary emission from $\tau$~Boo might not always be on. This would imply that the magnetosphere dynamics of $\tau$~Boo~b would be different than that of Jupiter. For example, the density of the magnetosphere could be variable and perhaps temporarily become void of particles.}
\end{enumerate}

A large follow-up campaign is needed to disentangle the true cause of the possible variability of the planetary radio emission. Multi-site observations are highly encouraged to rule out instrumental systematics. Monitoring the planet throughout its orbit and the stellar magnetic cycle is necessary to disentangle the competing effects. There is also a need for measurements of the stellar magnetic field taken throughout the radio follow-up campaign. Such a coordinated campaign is currently ongoing, using NenuFAR for the radio emission, and the TBL/Neo-Narval to monitor the magnetic field of the host star.

With our non-detection, we can place an upper limit on the radio emission from the $\tau$~Boo system at the time of the observations. Since NenuFAR is still in early commissioning the system equivalent flux density (SEFD) is currently not well characterized. Therefore, we will approximate our upper limit of NenuFAR based-off the sensitivity limit found for LOFAR in \citet{Turner2019}. The theoretical sensitivity limit of LOFAR for broadband bursts 
\begin{align}
    \sigma_{LOFAR} =& \frac{S_{Sys}}{N\sqrt{b\tau}}, \label{eq:sigmaLOFAR_nonumbers}\\
    \sigma_{LOFAR} =& \frac{40 kJy}{24\sqrt{1 \ sec \times 18\ MHz}} = 393 \ mJy \label{eq:sigmaLOFAR} 
\end{align}
where $N$ is the number of stations (24), $b$ is the bandwidth (18 MHz using the 15-33 MHz frequency range), $\tau$ is the timing resolution (1 sec), and $S_{Sys}$ is the station SEFD with a value of 40 kJy (\citealt{vanHaarlem2013}). \citet{Turner2019} measured the sensitivity limit of LOFAR to be $1.3\times\sigma_{LOFAR}$ and we will use this value below. For the burst emission from NenuFAR, we find a 3$\sigma$ flux upper limit of 1.5 Jy and 590 mJy assuming NenuFAR is as sensitive and twice as sensitive as LOFAR, respectively. This limit is a conservative preliminary estimate as NenuFAR is expected to be more sensitive (at least 2--5$\times$) than LOFAR between 15--30 MHz. In the future, the SEFD of NenuFAR will be accurately measured and we will perform a comprehensive sensitivity estimation using Jupiter radio emission simulated as if it was an exoplanet (using the same method used to obtain the sensitivity limit for LOFAR observations in \citealt{Turner2019}). Therefore, we expect to have a more precise upper limit in future work. 

\correction{The current NenuFAR observations are informative on the nature of $\tau$~Boo~b's radio emission. The original tentative detection of burst emission from \citetalias{Turner2021} was observed to have a flux density of 890 mJy. Therefore, we would have detected the burst emission of $\tau$~Boo~b (in the absence of variability, see above) at $\sim$4.5 $\sigma$ using the NenuFAR conservative sensitivity limit (2$\times$ as sensitive as LOFAR) of $\sim$200 mJy. We also expected to observe bursty emission at an orbital phase (e.g. 0.8) similar to that found in \citetalias{Turner2021} and validated in the model by \citet{Ashtari2022}. Therefore, the observed phases and the current upper limits can be used as priors in theoretical predictions (e.g \citealt{Elekes2023,Ashtari2022}) in the expected variability of $\tau$~Boo~b's radio emission. } 

\section{Conclusions and Perspectives}
In this study, we search for bursty emission from the $\tau$~Boo exoplanetary system using NenuFAR during its commissioning phase in 2020. \correction{This is the first exoplanet observation from NenuFAR to be published.} These observations are part of a large ongoing follow-up campaign to confirm the tentative detections of radio emission from $\tau$~Boo observed by LOFAR (\citealt{Turner2021}). We do not detect any bursty emission in the NenuFAR observations and the reason for this non-detection is degenerate. Possible interpretations include
the possibility that the original burst detection was an unknown systematic error or that radio emission from $\tau$~Boo~b is variable in time. Further analysis on the NenuFAR data and more observations are needed to fully understand the radio emission from the $\tau$~Boo system. Such observations are currently ongoing.

\section*{Acknowledgements} 

J.D. Turner was supported for this work by NASA through the NASA Hubble Fellowship grant $\#$HST-HF2-51495.001-A awarded by the Space Telescope Science Institute, which is operated by the Association of Universities for Research in Astronomy, Incorporated, under NASA contract NAS5-26555.

This paper is based on data obtained using the NenuFAR radio-telescope. The development of NenuFAR has been supported by personnel and funding from: Observatoire Radioastronomique de Nan\c{c}ay, CNRS-INSU, Observatoire de Paris-PSL, Universit\'{e} d'Orl\'{e}ans, Observatoire des Sciences de l'Univers en R\'{e}gion Centre, R\'{e}gion Centre-Val de Loire, DIM-ACAV and DIM-ACAV+ of R\'{e}gion Ile-de-France, Agence Nationale de la Recherche.

We acknowledge the use of the Nan\c{c}ay Data Center computing facility (CDN - Centre de Donn\'{e}es de Nan\c{c}ay). The CDN is hosted by the Station de Radioastronomie de Nan\c{c}ay in partnership with Observatoire de Paris, Universit\'{e} d'Orl\'{e}ans, OSUC and the CNRS. The CDN is supported by the Region Centre-Val de Loire, D\'{e}partement du Cher.

This work was supported by the Programme National de Plan\'{e}tologie (PNP) of CNRS/INSU co-funded by CNES and by the Programme National de Physique Stellaire (PNPS) of CNRS/INSU co-funded by CEA and CNES. PZ acknowledges funding from the ERC under the European Union's Horizon 2020 research and innovation programme (grant agreement no. 101020459 - Exoradio).

The Nan\c{c}ay Radio Observatory is operated by the Paris Observatory, associated with the French Centre National de la Recherche Scientifique (CNRS).

This research has made use of the Extrasolar Planet Encyclopaedia (exoplanet.eu) maintained by J. Schneider (\citealt{Schneider2011}), the NASA Exoplanet Archive, which is operated by the California Institute of Technology, under contract with the National Aeronautics and Space Administration under the Exoplanet Exploration Program, and NASA's Astrophysics Data System Bibliographic Services. In this paper, all the physical characteristics for the pulsar B0809+74 were taken from the ATNF Pulsar Catalogue \citep{Manchester2005} located at http://www.atnf.csiro.au/research/pulsar/psrcat. 

\correction{We thank the anonymous referees for their useful and thoughtful comments.}

\newcommand{\newblock}{}
\bibliographystyle{mnras}
\bibliography{bibliography.bib}

\begin{thebibliography}{}
\makeatletter
\relax
\def\mn@urlcharsother{\let\do\@makeother \do\$\do\&\do\#\do\^\do\_\do\%\do\~}
\def\mn@doi{\begingroup\mn@urlcharsother \@ifnextchar [ {\mn@doi@}
  {\mn@doi@[]}}
\def\mn@doi@[#1]#2{\def\@tempa{#1}\ifx\@tempa\@empty \href
  {http://dx.doi.org/#2} {doi:#2}\else \href {http://dx.doi.org/#2} {#1}\fi
  \endgroup}
\def\mn@eprint#1#2{\mn@eprint@#1:#2::\@nil}
\def\mn@eprint@arXiv#1{\href {http://arxiv.org/abs/#1} {{\tt arXiv:#1}}}
\def\mn@eprint@dblp#1{\href {http://dblp.uni-trier.de/rec/bibtex/#1.xml}
  {dblp:#1}}
\def\mn@eprint@#1:#2:#3:#4\@nil{\def\@tempa {#1}\def\@tempb {#2}\def\@tempc
  {#3}\ifx \@tempc \@empty \let \@tempc \@tempb \let \@tempb \@tempa \fi \ifx
  \@tempb \@empty \def\@tempb {arXiv}\fi \@ifundefined
  {mn@eprint@\@tempb}{\@tempb:\@tempc}{\expandafter \expandafter \csname
  mn@eprint@\@tempb\endcsname \expandafter{\@tempc}}}

\bibitem[\protect\citeauthoryear{{Alexander}, {Wynn}, {Mohammed}, {Nichols}  \&
  {Ercolano}}{{Alexander} et~al.}{2016}]{Alexander2016}
{Alexander} R.~D.,  {Wynn} G.~A.,  {Mohammed} H.,  {Nichols} J.~D.,
  {Ercolano} B.,  2016, {Magnetospheres of hot Jupiters: hydrodynamic models
  and ultraviolet absorption}, \textit{\mn@doi [MNRAS]
  {10.1093/mnras/stv2867}}, \href
  {https://ui.adsabs.harvard.edu/abs/2016MNRAS.456.2766A} {\textit{456, 2766}}

\bibitem[\protect\citeauthoryear{{Ashtari}, {Sciola}, {Turner}  \&
  {Stevenson}}{{Ashtari} et~al.}{2022}]{Ashtari2022}
{Ashtari} R.,  {Sciola} A.,  {Turner} J.~D.,   {Stevenson} K.,  2022,
  {Detecting Magnetospheric Radio Emission from Giant Exoplanets},
  \textit{\mn@doi [ApJ] {10.3847/1538-4357/ac92f5}}, \href
  {https://ui.adsabs.harvard.edu/abs/2022ApJ...939...24A} {\textit{939, 24}}

\bibitem[\protect\citeauthoryear{{Bastian}, {Dulk}  \& {Leblanc}}{{Bastian}
  et~al.}{2000}]{Bastian2000}
{Bastian} T.~S.,  {Dulk} G.~A.,   {Leblanc} Y.,  2000, {A Search for Radio
  Emission from Extrasolar Planets}, \textit{\mn@doi [ApJ] {10.1086/317864}},
  \href {http://adsabs.harvard.edu/abs/2000ApJ...545.1058B} {\textit{545,
  1058}}

\bibitem[\protect\citeauthoryear{{Bastian}, {Villadsen}, {Maps}, {Hallinan}  \&
  {Beasley}}{{Bastian} et~al.}{2018}]{Bastian2018}
{Bastian} T.~S.,  {Villadsen} J.,  {Maps} A.,  {Hallinan} G.,   {Beasley}
  A.~J.,  2018, {Radio Emission from the Exoplanetary System {\ensuremath{\in}}
  Eridani}, \textit{\mn@doi [ApJ] {10.3847/1538-4357/aab3cb}}, \href
  {https://ui.adsabs.harvard.edu/abs/2018ApJ...857..133B} {\textit{857, 133}}

\bibitem[\protect\citeauthoryear{{Blanco-Pozo} et~al.,}{{Blanco-Pozo}
  et~al.}{2023}]{BlancoPozo2023}
{Blanco-Pozo} J.,  et~al., 2023, {The CARMENES search for exoplanets around M
  dwarfs. A long-period planet around GJ 1151 measured with CARMENES and
  HARPS-N data}, \textit{\mn@doi [A$\&$A] {10.1051/0004-6361/202245053}}, \href{https://arxiv.org/abs/2301.04442}{\textit{671, A50}}

\bibitem[\protect\citeauthoryear{{Bondonneau} et~al.,}{{Bondonneau}
  et~al.}{2021}]{Bondonneau2021}
{Bondonneau} L.,  et~al., 2021, {Pulsars with NenuFAR: Backend and pipelines},
  \textit{\mn@doi [A$\&$A] {10.1051/0004-6361/202039339}}, \href{https://ui.adsabs.harvard.edu/abs/2021A&A...652A..34B} {\textit{652, A34}}

\bibitem[\protect\citeauthoryear{{Callingham} et~al.,}{{Callingham}
  et~al.}{2021}]{Callingham2021}
{Callingham} J.~R.,  et~al., 2021, {The population of M dwarfs observed at low
  radio frequencies}, \textit{\mn@doi [Nature Astronomy]
  {10.1038/s41550-021-01483-0}}, \href
  {https://ui.adsabs.harvard.edu/abs/2021NatAs...5.1233C} {\textit{5, 1233}}

\bibitem[\protect\citeauthoryear{{Carter} \& {Winn}}{{Carter} \&
  {Winn}}{2009}]{Carter2009}
{Carter} J.~A.,  {Winn} J.~N.,  2009, {Parameter Estimation from Time-series
  Data with Correlated Errors: A Wavelet-based Method and its Application to
  Transit Light Curves}, \textit{\mn@doi [ApJ] {10.1088/0004-637X/704/1/51}},
  \href {https://ui.adsabs.harvard.edu/abs/2009ApJ...704...51C} {\textit{704,
  51}}

\bibitem[\protect\citeauthoryear{{Cauley}, {Shkolnik}, {Llama}  \&
  {Lanza}}{{Cauley} et~al.}{2019}]{Cauley2019}
{Cauley} P.~W.,  {Shkolnik} E.~L.,  {Llama} J.,   {Lanza} A.~F.,  2019,
  {Magnetic field strengths of hot Jupiters from signals of star-planet
  interactions}, \textit{\mn@doi [Nature Astronomy]
  {10.1038/s41550-019-0840-x}}, \href
  {https://ui.adsabs.harvard.edu/abs/2019NatAs.tmp..408C} {\textit{p.~408}}

\bibitem[\protect\citeauthoryear{{Cendes}, {Williams}  \& {Berger}}{{Cendes}
  et~al.}{2022}]{Cendes2022}
{Cendes} Y.,  {Williams} P.~K.~G.,   {Berger} E.,  2022, {A Pilot Radio Search
  for Magnetic Activity in Directly Imaged Exoplanets}, \textit{\mn@doi [AJ]
  {10.3847/1538-3881/ac32c8}}, \href
  {https://ui.adsabs.harvard.edu/abs/2022AJ....163...15C} {\textit{163, 15}}

\bibitem[\protect\citeauthoryear{{Elekes} \& {Saur}}{{Elekes} \&
  {Saur}}{2023}]{Elekes2023}
{Elekes} F.,  {Saur} J.,  2023, {Space environment and magnetospheric Poynting
  fluxes of the exoplanet $\tau$ Bo{\"o}tis b}, \textit{\mn@doi [arXiv
  e-prints] {10.48550/arXiv.2301.05015}}, \href
  {https://ui.adsabs.harvard.edu/abs/2023arXiv230105015E} {\textit{p.
  arXiv:2301.05015}}

\bibitem[\protect\citeauthoryear{{Fares} et~al.,}{{Fares}
  et~al.}{2009}]{Fares2009}
{Fares} R.,  et~al., 2009, {Magnetic cycles of the planet-hosting star
  {\ensuremath{\tau}} Bootis - II. A second magnetic polarity reversal},
  \textit{\mn@doi [MNRAS] {10.1111/j.1365-2966.2009.15303.x}}, \href
  {https://ui.adsabs.harvard.edu/abs/2009MNRAS.398.1383F} {\textit{398, 1383}}

\bibitem[\protect\citeauthoryear{{Fares} et~al.,}{{Fares}
  et~al.}{2010}]{Fares2010}
{Fares} R.,  et~al., 2010, {Searching for star-planet interactions within the
  magnetosphere of HD189733}, \textit{\mn@doi [MNRAS]
  {10.1111/j.1365-2966.2010.16715.x}}, \href
  {https://ui.adsabs.harvard.edu/abs/2010MNRAS.406..409F} {\textit{406, 409}}

\bibitem[\protect\citeauthoryear{{Farrell}, {Desch}  \& {Zarka}}{{Farrell}
  et~al.}{1999}]{Farrell1999}
{Farrell} W.~M.,  {Desch} M.~D.,   {Zarka} P.,  1999, {On the possibility of
  coherent cyclotron emission from extrasolar planets}, \textit{\mn@doi [J.
  Geophys. Res.] {10.1029/1998JE900050}}, \href
  {http://adsabs.harvard.edu/abs/1999JGR...10414025F} {\textit{104, 14025}}

\bibitem[\protect\citeauthoryear{{Grie{\ss}meier}}{{Grie{\ss}meier}}{2015}]{G2015}
{Grie{\ss}meier} J.-M.,  2015, {Detection Methods and Relevance of Exoplanetary
  Magnetic Fields}, \textit{in Astrophysics and Space Science Library,  eds
  {Lammer} H.,  {Khodachenko} M., ,  Astrophysics and Space Science Library
  Vol. 411}, p.~213, \mn@doi{10.1007/978-3-319-09749-7_11}

\bibitem[\protect\citeauthoryear{Grie{\ss}meier}{Grie{\ss}meier}{2017}]{Griessmeier17PREVIII}
Grie{\ss}meier J.-M.,  2017, The search for radio emission from giant
  exoplanets, \textit{in Planetary Radio Emissions VIII,  eds G. Fischer and G.
  Mann and M. Panchenko and P. Zarka}, Austrian Academy of Sciences Press,
  Vienna, pp 285--300

\bibitem[\protect\citeauthoryear{{Grie{\ss}meier}}{{Grie{\ss}meier}}{2018}]{Griessmeier2018haex}
{Grie{\ss}meier} J.~M.,  2018, {Future Exoplanet Research: Radio Detection and
  Characterization}, \textit{in Handbook of Exoplanets, ISBN 978-3-319-55332-0.
  Springer International Publishing AG, part of Springer Nature, 2018, id.159},
  p.~159, \mn@doi{10.1007/978-3-319-55333-7_159}

\bibitem[\protect\citeauthoryear{{Grie{\ss}meier}, {Stadelmann}, {Motschmann},
  {Belisheva}, {Lammer}  \& {Biernat}}{{Grie{\ss}meier} et~al.}{2005a}]{Gr2005}
{Grie{\ss}meier} J.-M.,  {Stadelmann} A.,  {Motschmann} U.,  {Belisheva} N.~K.,
   {Lammer} H.,   {Biernat} H.~K.,  2005a, {Cosmic Ray Impact on Extrasolar
  Earth-Like Planets in Close-in Habitable Zones}, \textit{\mn@doi
  [Astrobiology] {10.1089/ast.2005.5.587}}, \href
  {http://adsabs.harvard.edu/abs/2005AsBio...5..587G} {\textit{5, 587}}

\bibitem[\protect\citeauthoryear{Grie{\ss}meier, Motschmann, Mann  \&
  Rucker}{Grie{\ss}meier et~al.}{2005b}]{Griessmeier05AA}
Grie{\ss}meier J.-M.,  Motschmann U.,  Mann G.,   Rucker H.~O.,  2005b, The
  influence of stellar wind conditions on the detectability of planetary radio
  emissions, \textit{A$\&$A}, \textit{437, 717}

\bibitem[\protect\citeauthoryear{{Grie{\ss}meier}, {Preusse}, {Khodachenko},
  {Motschmann}, {Mann}  \& {Rucker}}{{Grie{\ss}meier}
  et~al.}{2007a}]{Griessmeier07PSS}
{Grie{\ss}meier} J.~M.,  {Preusse} S.,  {Khodachenko} M.,  {Motschmann} U.,
  {Mann} G.,   {Rucker} H.~O.,  2007a, {Exoplanetary radio emission under
  different stellar wind conditions}, \textit{\mn@doi [Planetary and Space
  Science] {10.1016/j.pss.2006.01.008}}, \href
  {https://ui.adsabs.harvard.edu/abs/2007P&SS...55..618G} {\textit{55, 618}}

\bibitem[\protect\citeauthoryear{{Grie{\ss}meier}, {Zarka}  \&
  {Spreeuw}}{{Grie{\ss}meier} et~al.}{2007b}]{Gr2007}
{Grie{\ss}meier} J.-M.,  {Zarka} P.,   {Spreeuw} H.,  2007b, {Predicting
  low-frequency radio fluxes of known extrasolar planets}, \textit{\mn@doi
  [A$\&$A] {10.1051/0004-6361:20077397}}, \href
  {http://adsabs.harvard.edu/abs/2007A%26A...475..359G} {\textit{475, 359}}

\bibitem[\protect\citeauthoryear{{Grie{\ss}meier}, {Erkaev}, {Weber}, Lammer,
  Ivano  \& Odert}{{Grie{\ss}meier} et~al.}{2023}]{Griessmeier23PRE9}
{Grie{\ss}meier} J.-M.,  {Erkaev} N.~V.,  {Weber} C.,  Lammer H.,  Ivano V.~A.,
    Odert P.,  2023, {Required conditions for an exoplanet to emit radio waves
  and implications for observational campaigns}, \textit{in Planetary Radio
  Emissions IX,  ed. C. ~K.~Louis, C.~M.~Jackman, G.~Fischer, A.~H. ~Sulaiman,
  P.~Zucca}, Austrian Academy of Sciences Press, Vienna,
  \mn@doi{https://doi.org/10.25546/103090}

\bibitem[\protect\citeauthoryear{{Hallinan}, {Sirothia}, {Antonova},
  {Ishwara-Chandra}, {Bourke}, {Doyle}, {Hartman}  \& {Golden}}{{Hallinan}
  et~al.}{2013}]{Hallinan2013}
{Hallinan} G.,  {Sirothia} S.~K.,  {Antonova} A.,  {Ishwara-Chandra} C.~H.,
  {Bourke} S.,  {Doyle} J.~G.,  {Hartman} J.,   {Golden} A.,  2013, {Looking
  for a Pulse: A Search for Rotationally Modulated Radio Emission from the Hot
  Jupiter, {$\tau$} Bo{\"o}tis b}, \textit{\mn@doi [ApJ]
  {10.1088/0004-637X/762/1/34}}, \href
  {http://adsabs.harvard.edu/abs/2013ApJ...762...34H} {\textit{762, 34}}

\bibitem[\protect\citeauthoryear{{Heald} et~al.,}{{Heald}
  et~al.}{2015}]{Heald2015}
{Heald} G.~H.,  et~al., 2015, {The LOFAR Multifrequency Snapshot Sky Survey
  (MSSS). I. Survey description and first results}, \textit{\mn@doi [A$\&$A]
  {10.1051/0004-6361/201425210}}, \href
  {https://ui.adsabs.harvard.edu/abs/2015A%26A...582A.123H} {\textit{582,
  A123}}

\bibitem[\protect\citeauthoryear{{Hess} \& {Zarka}}{{Hess} \&
  {Zarka}}{2011}]{Hess2011}
{Hess} S.~L.~G.,  {Zarka} P.,  2011, {Modeling the radio signature of the
  orbital parameters, rotation, and magnetic field of exoplanets},
  \textit{\mn@doi [A$\&$A] {10.1051/0004-6361/201116510}}, \href
  {http://adsabs.harvard.edu/abs/2011A%26A...531A..29H} {\textit{531, A29}}

\bibitem[\protect\citeauthoryear{{Hindle}, {Bushby}  \& {Rogers}}{{Hindle}
  et~al.}{2021}]{Hindle2021}
{Hindle} A.~W.,  {Bushby} P.~J.,   {Rogers} T.~M.,  2021, {The Magnetic
  Mechanism for Hotspot Reversals in Hot Jupiter Atmospheres}, \textit{\mn@doi
  [ApJ] {10.3847/1538-4357/ac0e2e}}, \href
  {https://ui.adsabs.harvard.edu/abs/2021ApJ...922..176H} {\textit{922, 176}}

\bibitem[\protect\citeauthoryear{{Jeffers} et~al.,}{{Jeffers}
  et~al.}{2018}]{Jeffers2018}
{Jeffers} S.~V.,  et~al., 2018, {The relation between stellar magnetic field
  geometry and chromospheric activity cycles - II The rapid 120-day magnetic
  cycle of {\ensuremath{\tau}} Bootis}, \textit{\mn@doi [MNRAS]
  {10.1093/mnras/sty1717}}, \href
  {https://ui.adsabs.harvard.edu/abs/2018MNRAS.479.5266J} {\textit{479, 5266}}

\bibitem[\protect\citeauthoryear{{Kao}, {Hallinan}, {Pineda}, {Escala},
  {Burgasser}, {Bourke}  \& {Stevenson}}{{Kao} et~al.}{2016}]{Kao2016}
{Kao} M.~M.,  {Hallinan} G.,  {Pineda} J.~S.,  {Escala} I.,  {Burgasser} A.,
  {Bourke} S.,   {Stevenson} D.,  2016, {Auroral Radio Emission from Late L and
  T Dwarfs: A New Constraint on Dynamo Theory in the Substellar Regime},
  \textit{\mn@doi [ApJ] {10.3847/0004-637X/818/1/24}}, \href
  {https://ui.adsabs.harvard.edu/abs/2016ApJ...818...24K} {\textit{818, 24}}

\bibitem[\protect\citeauthoryear{{Kao}, {Hallinan}, {Pineda}, {Stevenson}  \&
  {Burgasser}}{{Kao} et~al.}{2018}]{Kao2018}
{Kao} M.~M.,  {Hallinan} G.,  {Pineda} J.~S.,  {Stevenson} D.,   {Burgasser}
  A.,  2018, {The Strongest Magnetic Fields on the Coolest Brown Dwarfs},
  \textit{\mn@doi [The Astrophysical Journal Supplement Series]
  {10.3847/1538-4365/aac2d5}}, \href
  {https://ui.adsabs.harvard.edu/abs/2018ApJS..237...25K} {\textit{237, 25}}

\bibitem[\protect\citeauthoryear{{Knierim}, {Batygin}  \& {Bitsch}}{{Knierim}
  et~al.}{2022}]{Knierim2022}
{Knierim} H.,  {Batygin} K.,   {Bitsch} B.,  2022, {Shallowness of circulation
  in hot Jupiters. Advancing the Ohmic dissipation model}, \textit{\mn@doi
  [A$\&$A] {10.1051/0004-6361/202142588}}, \href
  {https://ui.adsabs.harvard.edu/abs/2022A&A...658L...7K} {\textit{658, L7}}

\bibitem[\protect\citeauthoryear{{Lammer} et~al.,}{{Lammer}
  et~al.}{2009}]{Lammer2009}
{Lammer} H.,  et~al., 2009, {What makes a planet habitable?}, \textit{\mn@doi
  [A\&Ar] {10.1007/s00159-009-0019-z}}, \href
  {http://adsabs.harvard.edu/abs/2009A%26ARv..17..181L} {\textit{17, 181}}

\bibitem[\protect\citeauthoryear{{Lazio}, {Farrell}, {Dietrick}, {Greenlees},
  {Hogan}, {Jones}  \& {Hennig}}{{Lazio} et~al.}{2004}]{Lazio2004}
{Lazio} W. T.~J.,  {Farrell} W.~M.,  {Dietrick} J.,  {Greenlees} E.,  {Hogan}
  E.,  {Jones} C.,   {Hennig} L.~A.,  2004, {The Radiometric Bode's Law and
  Extrasolar Planets}, \textit{\mn@doi [ApJ] {10.1086/422449}}, \href
  {http://adsabs.harvard.edu/abs/2004ApJ...612..511L} {\textit{612, 511}}

\bibitem[\protect\citeauthoryear{{Lazio}, {Shkolnik}, {Hallinan}  \& {Planetary
  Habitability Study Team}}{{Lazio} et~al.}{2016}]{Lazio2016}
{Lazio} T.~J.~W.,  {Shkolnik} E.,  {Hallinan} G.,   {Planetary Habitability
  Study Team} 2016, Technical report, {Planetary Magnetic Fields: Planetary
  Interiors and Habitability}

\bibitem[\protect\citeauthoryear{{Lazio} et~al.,}{{Lazio}
  et~al.}{2019}]{Lazio2019}
{Lazio} J.,  et~al., 2019, {Magnetic Fields of Extrasolar Planets: Planetary
  Interiors and Habitability}, \textit{BAAS}, \href
  {https://ui.adsabs.harvard.edu/abs/2019BAAS...51c.135L} {\textit{51, 135}}

\bibitem[\protect\citeauthoryear{{Lecavelier des Etangs}, {Sirothia},
  {Gopal-Krishna}  \& {Zarka}}{{Lecavelier des Etangs}
  et~al.}{2013}]{Lecavelier2013}
{Lecavelier des Etangs} A.,  {Sirothia} S.~K.,  {Gopal-Krishna}  {Zarka} P.,
  2013, {Hint of 150 MHz radio emission from the Neptune-mass extrasolar
  transiting planet HAT-P-11b}, \textit{\mn@doi [A\&A]
  {10.1051/0004-6361/201219789}}, \href
  {http://adsabs.harvard.edu/abs/2013A%26A...552A..65L} {\textit{552, A65}}

\bibitem[\protect\citeauthoryear{{Lynch}, {Murphy}, {Lenc}  \&
  {Kaplan}}{{Lynch} et~al.}{2018}]{Lynch2018}
{Lynch} C.~R.,  {Murphy} T.,  {Lenc} E.,   {Kaplan} D.~L.,  2018, {The
  detectability of radio emission from exoplanets}, \textit{\mn@doi [MNRAS]
  {10.1093/mnras/sty1138}}, \href
  {http://adsabs.harvard.edu/abs/2018MNRAS.478.1763L} {\textit{478, 1763}}

\bibitem[\protect\citeauthoryear{{Manchester}, {Hobbs}, {Teoh}  \&
  {Hobbs}}{{Manchester} et~al.}{2005}]{Manchester2005}
{Manchester} R.~N.,  {Hobbs} G.~B.,  {Teoh} A.,   {Hobbs} M.,  2005, {The
  Australia Telescope National Facility Pulsar Catalogue}, \textit{\mn@doi [AJ]
  {10.1086/428488}}, \href
  {https://ui.adsabs.harvard.edu/abs/2005AJ....129.1993M} {\textit{129, 1993}}

\bibitem[\protect\citeauthoryear{{Mauduit}, {Zarka}, {Grie{\ss}meier}  \&
  {Turner}}{{Mauduit} et~al.}{2023}]{Mauduit23PRE9}
{Mauduit} E.,  {Zarka} P.,  {Grie{\ss}meier} J.-M.,   {Turner} J.~D.,  2023,
  {PALANTIR: an updated prediction tool for exoplanetary radioemissions},
  \textit{in Planetary Radio Emissions IX,  ed. C. ~K.~Louis, C.~M.~Jackman,
  G.~Fischer, A.~H. ~Sulaiman, P.~Zucca}, Austrian Academy of Sciences Press,
  Vienna, \mn@doi{https://doi.org/10.25546/103092}

\bibitem[\protect\citeauthoryear{{McIntyre}, {Lineweaver}  \& {Ireland
  }}{{McIntyre} et~al.}{2019}]{McIntyre2019}
{McIntyre} S. R.~N.,  {Lineweaver} C.~H.,   {Ireland } M.~J.,  2019, {Planetary
  magnetism as a parameter in exoplanet habitability}, \textit{\mn@doi [MNRAS]
  {10.1093/mnras/stz667}}, \href
  {https://ui.adsabs.harvard.edu/abs/2019MNRAS.485.3999M} {\textit{485, 3999}}

\bibitem[\protect\citeauthoryear{{Meadows} \& {Barnes}}{{Meadows} \&
  {Barnes}}{2018}]{Meadows2018haex}
{Meadows} V.~S.,  {Barnes} R.~K.,  2018, in  eds {Deeg} H.~J.,  {Belmonte}
  J.~A., , , Handbook of Exoplanets.
p.~57, \mn@doi{10.1007/978-3-319-55333-7\_57}

\bibitem[\protect\citeauthoryear{{Narang}, {Oza}, {Hakim}, {Manoj}, {Banyal}
  \& {Thorngren}}{{Narang} et~al.}{2023}]{Narang2023}
{Narang} M.,  {Oza} A.~V.,  {Hakim} K.,  {Manoj} P.,  {Banyal} R.~K.,
  {Thorngren} D.~P.,  2023, {Radio-loud Exoplanet-exomoon Survey: GMRT Search
  for Electron Cyclotron Maser Emission}, \textit{\mn@doi [AJ]
  {10.3847/1538-3881/ac9eb8}}, \href
  {https://ui.adsabs.harvard.edu/abs/2023AJ....165....1N} {\textit{165, 1}}

\bibitem[\protect\citeauthoryear{{Nichols}}{{Nichols}}{2012}]{Nichols2012}
{Nichols} J.~D.,  2012, {Candidates for detecting exoplanetary radio emissions
  generated by magnetosphere-ionosphere coupling}, \textit{\mn@doi [MNRAS]
  {10.1111/j.1745-3933.2012.01348.x}}, \href
  {http://adsabs.harvard.edu/abs/2012MNRAS.427L..75N} {\textit{427, L75}}

\bibitem[\protect\citeauthoryear{{P{\'e}rez-Torres} et~al.,}{{P{\'e}rez-Torres}
  et~al.}{2021}]{Perez2021}
{P{\'e}rez-Torres} M.,  et~al., 2021, {Monitoring the radio emission of Proxima
  Centauri}, \textit{\mn@doi [A$\&$A] {10.1051/0004-6361/202039052}}, \href{https://ui.adsabs.harvard.edu/abs/2021A&A...645A..77P} {\textit{645, A77}}

\bibitem[\protect\citeauthoryear{{Perna}, {Menou}  \& {Rauscher}}{{Perna}
  et~al.}{2010}]{Perna2010a}
{Perna} R.,  {Menou} K.,   {Rauscher} E.,  2010, {Magnetic Drag on Hot Jupiter
  Atmospheric Winds}, \textit{\mn@doi [ApJ] {10.1088/0004-637X/719/2/1421}},
  \href {Magnetic Drag on Hot Jupiter Atmospheric Winds} {\textit{719, 1421}}

\bibitem[\protect\citeauthoryear{{Pineda} \& {Villadsen}}{{Pineda} \&
  {Villadsen}}{2023}]{Pineda2023}
{Pineda} J.~S.,  {Villadsen} J.,  2023, {Coherent radio bursts from known
  M-dwarf planet-host YZ Ceti}, \textit{\mn@doi [Nature Astronomy]
  {10.1038/s41550-023-01914-0}}, \href
  {https://ui.adsabs.harvard.edu/abs/2023NatAs.tmp...65P} {\textit{}}

\bibitem[\protect\citeauthoryear{{Rauscher} \& {Menou}}{{Rauscher} \&
  {Menou}}{2013}]{Rauscher2013}
{Rauscher} E.,  {Menou} K.,  2013, {Three-dimensional Atmospheric Circulation
  Models of HD 189733b and HD 209458b with Consistent Magnetic Drag and Ohmic
  Dissipation}, \textit{\mn@doi [APJ] {10.1088/0004-637X/764/1/103}}, \href
  {http://adsabs.harvard.edu/abs/2013ApJ...764..103R} {\textit{764, 103}}

\bibitem[\protect\citeauthoryear{{Reiners} \& {Christensen}}{{Reiners} \&
  {Christensen}}{2010}]{Reiners2010}
{Reiners} A.,  {Christensen} U.~R.,  2010, {A magnetic field evolution scenario
  for brown dwarfs and giant planets}, \textit{\mn@doi [A$\&$A]
  {10.1051/0004-6361/201014251}}, \href
  {https://ui-adsabs-harvard-edu.insu.bib.cnrs.fr/abs/2010A&A...522A..13R}
  {\textit{522, A13}}

\bibitem[\protect\citeauthoryear{{Route}}{{Route}}{2019}]{Route2019}
{Route} M.,  2019, {The Rise of ROME. I. A Multiwavelength Analysis of the
  Star-Planet Interaction in the HD 189733 System}, \textit{\mn@doi [ApJ]
  {10.3847/1538-4357/aafc25}}, \href
  {https://ui.adsabs.harvard.edu/abs/2019ApJ...872...79R} {\textit{872, 79}}

\bibitem[\protect\citeauthoryear{{Schneider}, {Dedieu}, {Le Sidaner}, {Savalle}
   \& {Zolotukhin}}{{Schneider} et~al.}{2011}]{Schneider2011}
{Schneider} J.,  {Dedieu} C.,  {Le Sidaner} P.,  {Savalle} R.,   {Zolotukhin}
  I.,  2011, {Defining and cataloging exoplanets: the exoplanet.eu database},
  \textit{\mn@doi [A\&A] {10.1051/0004-6361/201116713}}, \href
  {http://adsabs.harvard.edu/abs/2011A%26A...532A..79S} {\textit{532, A79}}

\bibitem[\protect\citeauthoryear{See, Jardine, Fares, Donati  \& Moutou}{See
  et~al.}{2015}]{See2015}
See V.,  Jardine M.,  Fares R.,  Donati J.-F.,   Moutou C.,  2015, Time-scales
  of close-in exoplanet radio emission variability, \textit{\mn@doi [MNRAS]
  {10.1093/mnras/stv896}}, \textit{450, 4323}

\bibitem[\protect\citeauthoryear{{Sirothia}, {Lecavelier des Etangs},
  {Gopal-Krishna}, {Kantharia}  \& {Ishwar-Chandra}}{{Sirothia}
  et~al.}{2014}]{Sirothia2014}
{Sirothia} S.~K.,  {Lecavelier des Etangs} A.,  {Gopal-Krishna} {Kantharia}
  N.~G.,   {Ishwar-Chandra} C.~H.,  2014, {Search for 150 MHz radio emission
  from extrasolar planets in the TIFR GMRT Sky Survey}, \textit{\mn@doi [A\&A]
  {10.1051/0004-6361/201321571}}, \href
  {http://adsabs.harvard.edu/abs/2014A%26A...562A.108S} {\textit{562, A108}}

\bibitem[\protect\citeauthoryear{{Strugarek} et~al.,}{{Strugarek}
  et~al.}{2022}]{Strugarek2022}
{Strugarek} A.,  et~al., 2022, {MOVES - V. Modelling star-planet magnetic
  interactions of HD 189733}, \textit{\mn@doi [MNRAS] {10.1093/mnras/stac778}},
  \href {https://ui.adsabs.harvard.edu/abs/2022MNRAS.512.4556S} {\textit{512,
  4556}}

\bibitem[\protect\citeauthoryear{{Treumann}}{{Treumann}}{2006}]{Treumann2006}
{Treumann} R.~A.,  2006, {The electron-cyclotron maser for astrophysical
  application}, \textit{\mn@doi [A$\&$Ar] {10.1007/s00159-006-0001-y}}, \href
  {http://adsabs.harvard.edu/abs/2006A%26ARv..13..229T} {\textit{13, 229}}

\bibitem[\protect\citeauthoryear{{Trigilio} et~al.,}{{Trigilio}
  et~al.}{2023}]{Trigilio2023}
{Trigilio} C.,  et~al., 2023, {Star-Planet Interaction at radio wavelengths in
  YZ Ceti: Inferring planetary magnetic field}, \textit{\mn@doi [arXiv
  e-prints] {10.48550/arXiv.2305.00809}}, \href
  {https://ui.adsabs.harvard.edu/abs/2023arXiv230500809T} {\textit{p.
  arXiv:2305.00809}}

\bibitem[\protect\citeauthoryear{{Turner}, {Christie}, {Arras}, {Johnson}  \&
  {Schmidt}}{{Turner} et~al.}{2016a}]{Turner2016a}
{Turner} J.~D.,  {Christie} D.,  {Arras} P.,  {Johnson} R.~E.,   {Schmidt} C.,
  2016a, {Investigation of the environment around close-in transiting
  exoplanets using CLOUDY}, \textit{\mn@doi [MNRAS] {10.1093/mnras/stw556}},
  \href {http://adsabs.harvard.edu/abs/2016MNRAS.458.3880T} {\textit{458,
  3880}}

\bibitem[\protect\citeauthoryear{{Turner} et~al.,}{{Turner}
  et~al.}{2016b}]{Turner2016b}
{Turner} J.~D.,  et~al., 2016b, {Ground-based near-UV observations of 15
  transiting exoplanets: constraints on their atmospheres and no evidence for
  asymmetrical transits}, \textit{\mn@doi [MNRAS] {10.1093/mnras/stw574}},
  \href{http://adsabs.harvard.edu/abs/2016MNRAS.459..789T}{\textit{459, 789}}

\bibitem[\protect\citeauthoryear{{Turner}, {Grie{\ss}meier}, {Zarka}  \&
  {Vasylieva}}{{Turner} et~al.}{2017}]{Turner2017pre8}
{Turner} J.~D.,  {Grie{\ss}meier} J.-M.,  {Zarka} P.,   {Vasylieva} I.,  2017,
  {The search for radio emission from exoplanets using LOFAR low-frequency
  beam-formed observations: Data pipeline and preliminary results for the 55
  Cnc system}, \textit{in Planetary Radio Emissions VIII,  eds G. Fischer and
  G. Mann and M. Panchenko and P. Zarka}, Austrian Academy of Sciences Press,
  Vienna, pp 301--313

\bibitem[\protect\citeauthoryear{{Turner}, {Grie{\ss}meier}, {Zarka}  \&
  {Vasylieva}}{{Turner} et~al.}{2019}]{Turner2019}
{Turner} J.~D.,  {Grie{\ss}meier} J.-M.,  {Zarka} P.,   {Vasylieva} I.,  2019,
  {The search for radio emission from exoplanets using LOFAR beam-formed
  observations: Jupiter as an exoplanet}, \textit{\mn@doi [A$\&$A]
  {10.1051/0004-6361/201832848}}, \href{https://ui.adsabs.harvard.edu/abs/2019A&A...624A..40T}{\textit{624, A40}}

\bibitem[\protect\citeauthoryear{{Turner} et~al.,}{{Turner}
  et~al.}{2021}]{Turner2021}
{Turner} J.~D.,  et~al., 2021, {The search for radio emission from the
  exoplanetary systems 55 Cancri, {\ensuremath{\upsilon}} Andromedae, and
  {\ensuremath{\tau}} Bo{\"o}tis using LOFAR beam-formed observations},
  \textit{\mn@doi [A$\&$A] {10.1051/0004-6361/201937201}}, \href
  {https://ui.adsabs.harvard.edu/abs/2021A&A...645A..59T} {\textit{645, A59}}

\bibitem[\protect\citeauthoryear{Vasylieva}{Vasylieva}{2015}]{Vasylieva2015}
Vasylieva I.,  2015, {Pulsars and transients survey, and exoplanet search at
  low-frequencies with the UTR-2 radio telescope: methods and first results},
  Theses, \textit{{Paris Observatory}}, \url
  {https://tel.archives-ouvertes.fr/tel-01246634}

\bibitem[\protect\citeauthoryear{{Vedantham} et~al.,}{{Vedantham}
  et~al.}{2020}]{Vedantham2020}
{Vedantham} H.~K.,  et~al., 2020, {Coherent radio emission from a quiescent red
  dwarf indicative of star-planet interaction}, \textit{\mn@doi [Nature
  Astronomy] {10.1038/s41550-020-1011-9}}, \href
  {https://ui.adsabs.harvard.edu/abs/2020NatAs.tmp...34V} {\textit{}}

\bibitem[\protect\citeauthoryear{{Vidotto}, {Fares}, {Jardine}, {Donati},
  {Opher}, {Moutou}, {Catala}  \& {Gombosi}}{{Vidotto}
  et~al.}{2012}]{Vidotto2012}
{Vidotto} A.~A.,  {Fares} R.,  {Jardine} M.,  {Donati} J.~F.,  {Opher} M.,
  {Moutou} C.,  {Catala} C.,   {Gombosi} T.~I.,  2012, {The stellar wind cycles
  and planetary radio emission of the {\ensuremath{\tau}} Boo system},
  \textit{\mn@doi [MNRAS] {10.1111/j.1365-2966.2012.21122.x}}, \href
  {https://ui.adsabs.harvard.edu/abs/2012MNRAS.423.3285V} {\textit{423, 3285}}

\bibitem[\protect\citeauthoryear{{Wang} \& {Ford}}{{Wang} \&
  {Ford}}{2011}]{Wang2011}
{Wang} J.,  {Ford} E.~B.,  2011, {On the eccentricity distribution of
  short-period single-planet systems}, \textit{\mn@doi [MNRAS]
  {10.1111/j.1365-2966.2011.19600.x}}, \href
  {https://ui.adsabs.harvard.edu/abs/2011MNRAS.418.1822W} {\textit{418, 1822}}

\bibitem[\protect\citeauthoryear{{Weber} et~al.,}{{Weber}
  et~al.}{2017}]{Weber2017pre8}
{Weber} C.,  et~al., 2017, {On the Cyclotron Maser Instability in
  Magnetospheres of Hot Jupiters - Influence of ionosphere models}, \textit{in
  Planetary Radio Emissions VIII,  eds G. Fischer and G. Mann and M. Panchenko
  and P. Zarka}, Austrian Academy of Sciences Press, Vienna, pp 317--329,
  \mn@doi{10.1553/PRE8s317}

\bibitem[\protect\citeauthoryear{{Weber}, {Erkaev}, {Ivanov}, {Odert},
  {Grie{\ss}meier}, {Fossati}, {Lammer}  \& {Rucker}}{{Weber}
  et~al.}{2018}]{Weber2018}
{Weber} C.,  {Erkaev} N.~V.,  {Ivanov} V.~A.,  {Odert} P.,  {Grie{\ss}meier}
  J.-M.,  {Fossati} L.,  {Lammer} H.,   {Rucker} H.~O.,  2018, {Supermassive
  hot Jupiters provide more favourable conditions for the generation of radio
  emission via the cyclotron maser instability - a case study based on Tau
  Bootis b}, \textit{\mn@doi [MNRAS] {10.1093/mnras/sty2079}}, \href
  {http://adsabs.harvard.edu/abs/2018MNRAS.480.3680W} {\textit{480, 3680}}

\bibitem[\protect\citeauthoryear{{Winglee}, {Dulk}  \& {Bastian}}{{Winglee}
  et~al.}{1986}]{Winglee1986}
{Winglee} R.~M.,  {Dulk} G.~A.,   {Bastian} T.~S.,  1986, {A search for
  cyclotron maser radiation from substellar and planet-like companions of
  nearby stars}, \textit{\mn@doi [ApJl] {10.1086/184760}}, \href
  {http://adsabs.harvard.edu/abs/1986ApJ...309L..59W} {\textit{309, L59}}

\bibitem[\protect\citeauthoryear{{Wu} \& {Lee}}{{Wu} \& {Lee}}{1979}]{Wu1979}
{Wu} C.~S.,  {Lee} L.~C.,  1979, {A theory of the terrestrial kilometric
  radiation}, \textit{\mn@doi [ApJ] {10.1086/157120}}, \href
  {http://adsabs.harvard.edu/abs/1979ApJ...230..621W} {\textit{230, 621}}

\bibitem[\protect\citeauthoryear{{Yantis}, {Sullivan}  \& {Erickson}}{{Yantis}
  et~al.}{1977}]{Yantis1977}
{Yantis} W.~F.,  {Sullivan} III W.~T.,   {Erickson} W.~C.,  1977, {A Search for
  Extra-Solar Jovian Planets by Radio Techniques}, \textit{in Bulletin of the
  American Astronomical Society}, p.~453

\bibitem[\protect\citeauthoryear{{Zarka}}{{Zarka}}{1998}]{Zarka1998}
{Zarka} P.,  1998, {Auroral radio emissions at the outer planets: Observations
  and theories}, \textit{\mn@doi [JGR] {10.1029/98JE01323}}, \href
  {http://adsabs.harvard.edu/abs/1998JGR...10320159Z} {\textit{103, 20159}}

\bibitem[\protect\citeauthoryear{{Zarka}}{{Zarka}}{2007}]{Zarka2007}
{Zarka} P.,  2007, {Plasma interactions of exoplanets with their parent star
  and associated radio emissions}, \textit{\mn@doi [Planetary and Space
  Science] {10.1016/j.pss.2006.05.045}}, \href
  {http://adsabs.harvard.edu/abs/2007P%26SS...55..598Z} {\textit{55, 598}}

\bibitem[\protect\citeauthoryear{{Zarka}}{{Zarka}}{2018}]{Zarka2018haex}
{Zarka} P.,  2018, {Star-Planet Interactions in the Radio Domain: Prospect for
  Their Detection}, \textit{in Handbook of Exoplanets}, p.~22,
  \mn@doi{10.1007/978-3-319-55333-7_22}

\bibitem[\protect\citeauthoryear{{Zarka} et~al.,}{{Zarka}
  et~al.}{1997}]{Zarka1997pre4}
{Zarka} P.,  et~al., 1997, {Ground-Based High Sensitivity Radio Astronomy at
  Decameter Wavelengths}, \textit{in Planetary Radio Emission IV,  eds
  {Rucker}, H.~O. and {Bauer}, S.~J. and {Lecacheux}, A.}, p.~101

\bibitem[\protect\citeauthoryear{{Zarka}, {Treumann}, {Ryabov}  \&
  {Ryabov}}{{Zarka} et~al.}{2001}]{Zarka2001}
{Zarka} P.,  {Treumann} R.~A.,  {Ryabov} B.~P.,   {Ryabov} V.~B.,  2001,
  {Magnetically-Driven Planetary Radio Emissions and Application to Extrasolar
  Planets}, \textit{\mn@doi [Astrophys. Space. Sci.]
  {10.1023/A:1012221527425}}, \href
  {http://adsabs.harvard.edu/abs/2001Ap%26SS.277..293Z} {\textit{277, 293}}

\bibitem[\protect\citeauthoryear{{Zarka}, {Cecconi}  \& {Kurth}}{{Zarka}
  et~al.}{2004}]{Zarka2004}
{Zarka} P.,  {Cecconi} B.,   {Kurth} W.~S.,  2004, {Jupiter's low-frequency
  radio spectrum from Cassini/Radio and Plasma Wave Science (RPWS) absolute
  flux density measurements}, \textit{\mn@doi [Journal of Geophysical Research
  (Space Physics)] {10.1029/2003JA010260}}, \href
  {http://adsabs.harvard.edu/abs/2004JGRA..109.9S15Z} {\textit{109, A09S15}}

\bibitem[\protect\citeauthoryear{{Zarka}, {Lazio}  \& {Hallinan}}{{Zarka}
  et~al.}{2015}]{Zarka2015SKA}
{Zarka} P.,  {Lazio} J.,   {Hallinan} G.,  2015, {Magnetospheric Radio
  Emissions from Exoplanets with the SKA}, \textit{Advancing Astrophysics with
  the Square Kilometre Array (AASKA14)}, \href
  {http://adsabs.harvard.edu/abs/2015aska.confE.120Z} {\textit{p.~120}}

\bibitem[\protect\citeauthoryear{{Zarka} et~al.}{{Zarka}
  et~al.}{2020}]{Zarka2020}
{Zarka} P.,  et~al., 2020, {The low-frequency radio telescope NenuFAR},
  \textit{in URSI General Assembly}, p. session J01: New Telescopes on the
  Frontier

\bibitem[\protect\citeauthoryear{{van Haarlem} et~al.,}{{van Haarlem}
  et~al.}{2013}]{vanHaarlem2013}
{van Haarlem} M.~P.,  et~al., 2013, {LOFAR: The LOw-Frequency ARray},
  \textit{\mn@doi [A$\&$A] {10.1051/0004-6361/201220873}}, \href
  {http://adsabs.harvard.edu/abs/2013A%26A...556A...2V} {\textit{556, A2}}

\makeatother
\end{thebibliography}

\end{document}